\documentstyle[sprocl]{article}
\input epsf
\newcommand{\CP}{\mbox{\it CP\,}}
\newcommand{\eg}{{\em e.g.\ }}

\newcommand{\vs}{{\em vs.\ }}
\begin{document}
\title 
{\begin{flushright}
{\vspace {-.05 in}
\normalsize IIT-HEP-97/1\\
hep-ex/9705002
}
\end{flushright}
\vskip 0.2in
\bf A Future Charm Facility\footnote{To appear in Proc.\ FCNC97 Symposium, 
Santa Monica, CA, 19--21 Feb.\ 1997.}}
\author{ D. M. Kaplan\footnote{E-mail: kaplan@fnal.gov}
     \\ {\sl Illinois Institute of Technology, Chicago, IL 60616, USA}
     \\
 for the BTeV Collaboration
 }  
\maketitle

\begin{abstract}
The ``BTeV/C0" experiment  at Fermilab could 
reconstruct $>$10$^9$ charm decays, four orders of magnitude beyond the 
largest extant sample. The experiment is likely to run during Tevatron 
Run II (ca.\ 2000--2005). In addition to ``programmatic" charm physics such as 
spectroscopy, lifetimes, and QCD tests, it will 
have significant new-physics reach in the areas of {\em CP} 
violation, flavor-changing neutral-current and lepton-number-violating decays, 
and $D^0\overline {D^0}$
mixing, and could observe direct {\em CP} violation in Cabibbo-suppressed 
$D$
decays if it occurs at the level predicted by the Standard Model.
\end{abstract}

\section{Introduction}

Charm sensitivities have increased exponentially over the last two
decades.
Current experiments aim to reconstruct $\sim10^6$ events, and the $B$ factories
and COMPASS facility\,\cite{COMPASS} could achieve $10^7$-event sensitivity. 
We are designing an experiment for the
Tevatron's C0 area which could reconstruct $10^9$ charm decays during Tevatron
Run II (ca.\ 2000--2005). While this ``BTeV/C0'' effort aims at both charm and
beauty physics, I focus here on charm.

Sensitivity at the proposed level will substantially advance such
``programmatic" charm physics as spectroscopy, lifetimes, and QCD tests. It
will also give substantial new-physics reach in the areas of {\em CP}
violation, flavor-changing neutral-current and lepton-number-violating decays,
and $D^0\overline {D^0}$ mixing. If direct {\em CP} violation occurs in
Cabibbo-suppressed $D$ decays at the level predicted by the Standard Model, it
could be observable in the BTeV/C0 experiment.

\section{Importance of Charm {\em CP}-Violation, Mixing, and Rare-Decay Studies}

{\CP} violation is recognized as one of the central problems of particle
physics. The mechanism(s) responsible for it have yet to be definitively
established. A leading candidate, the Kobayashi-Maskawa (KM) model,\,\cite{KM}
has the attractive feature of explaining the small size of $K^0$ {\CP}
asymmetries as a manifestation of the small mixing between the third
quark generation and the first two.\,\cite{Rosner,Wolfenstein-CP} 
Thus in the KM model, large {\CP} asymmetries
are expected in the beauty sector. Other models attribute the effect to the
exchange of massive particles such as $W$'s with right-handed coupings or extra
Higgs scalars.\,\cite{CP-survey} In these models {\CP} asymmetries should be
more ``democratic" and may be too small to observe in beauty 
($\cal{O}$$(10^{-3})$).
Many of these models predict large mixing in charm.

We do not know whether {\CP} violation arises exclusively from any one of these
mechanisms, whether many contribute, or whether some other mechanism not yet
thought of is the answer. Thus a balanced program of investigation in all
available quark (and lepton\,\cite{Tsai}) sectors is desirable. As is well
known, {\CP}-violation, mixing, and rare-decay studies in beauty are the goal
of several projects in progress around the world.
Such studies in charm are 
important precisely because the small Standard Model predictions can allow new 
physics to appear in a striking manner. 

\section{Charm {\em CP} Violation}

\subsection{Standard Model}

Direct {\em CP} violation in charm decay is expected in the Standard Model
(SM) at the $10^{-3}$ level\,\cite{CharmCP,Burdman} (see 
Table~\ref{tab:sens}). In the SM it is significant only for singly
Cabibbo-suppressed decays (SCSD), for which tree-level graphs can interfere
with penguin diagrams, leading to partial-decay-rate asymmetries:
\begin{equation} 
A \equiv \frac{\Gamma(D\to f)-\Gamma(\overline{D}\to\overline{f})}
{\Gamma(D\to f)+\Gamma(\overline{D}\to\overline{f})}
\neq 0\,, \end{equation}
where $\Gamma(D\to f)$ is the decay width for a $D$ meson to final state $f$
and 
$\Gamma(\overline{D}\to\overline{f})$ that for the {\em CP}-conjugate process.

{\footnotesize
\begin{table*}
\vspace{-0.1in}
\caption{Sensitivity to high-impact charm physics.$^*$ \label{tab:sens}}
\begin{center}
\begin{tabular}{|l|l|l|l|l|}
\hline
& & Reach of & SM \\ 
\raisebox{1.5ex}[0pt]{Topic} & \raisebox{1.5ex}[0pt]{Limit$\dagger$} &
``$10^8$-charm" exp't$\dagger$
& prediction \\
\hline\hline
Direct {\em CP} Viol.\ 
& & & \\ \hline
~$D^0\to K^- \pi^+$ & -0.009$<$$A$$<$0.027&  & $\approx0$ (CFD) 
\\
~$D^0\to K^- \pi^+\pi^-\pi^+$ & &  few$\,\times10^{-4}$ & $\approx0$ (CFD) 
\\
~$D^0\to K^+ \pi^-$ & 
& $10^{-3}-10^{-2}$ & $\approx0$ (DCSD) \\
~$D^+\to K^+ \pi^+ \pi^-$ & 
& few$\,\times10^{-3}$ & $\approx0$ (DCSD) \\
~$D^0\to K^- K^+$ & 
-0.11$<$$A$$<$0.16& $10^{-3}$ & 
$(0.13\pm0.8)\times10^{-3}$\\
& -0.028$<$$A$$<$0.166& & \\
~$D^+\to K^- K^+\pi^+$ & -0.062$<$$A$$<$0.034
& $10^{-3}$ & \\
~$D^+\to \overline {K^{*0}}K^+$ 
& -0.092$<$$A$$<$0.072& $10^{-3}$ & 
$(2.8\pm0.8)\times10^{-3}$\\
~$D^+\to \phi\pi^+$ & -0.087$<$$A$$<$0.031
& $10^{-3}$ & \\
~$D^+\to \pi^-\pi^+\pi^+$ & -0.086$<$$A$$<$0.052
& $10^{-3}$ & \\
~$D^+\to \rho^0\pi^+$ & & & $(-2.3\pm0.6)\times10^{-3}$\\
~$D^+\to \eta\pi^+$ & & & $(-1.5\pm0.4)\times10^{-3}$  \\
~$D^+\to K_S\pi^+$ & & few$\times10^{-4}$ & $3.3\times10^{-3}$\\
\hline
Indirect {\em CP} Viol.\ 
& & & \\ \hline
~$D^0\to\pi^+\pi^-$ & & few\,$\times10^{-3}$ & $\approx0$
\\
\hline
FCNC
& & & \\ \hline
~$D^0\to\mu^+\mu^-$ & $7.6\times 10^{-6}$ & $10^{-7}$ 
& $<3\times10^{-15}$ \\
~$D^0\to \pi^0\mu^+\mu^-$ & $1.7\times10^{-4}$ & $10^{-6}$ & \\
~$D^0\to \overline {K^0} e^+e^-$ & $17.0\times10^{-4}$
& $10^{-6}$ & $<2\times10^{-15}$ \\
~$D^0\to\overline {K^0}\mu^+\mu^-$ & $2.5\times10^{-4}$
& $10^{-6}$ & $<2\times10^{-15}$ \\
~$D^+\to \pi^+e^+e^-$ & $6.6\times10^{-5}$
& few$\,\times10^{-7}$ & $<10^{-8}$\\
~$D^+\to \pi^+\mu^+\mu^-$ & $1.8\times10^{-5}$
& few$\,\times10^{-7}$ & $<10^{-8}$\\
~$D^+\to K^+ e^+e^-$ & $4.8\times10^{-3}$
& few$\,\times10^{-7}$ & $<10^{-15}$\\
~$D^+\to K^+ \mu^+\mu^-$ & $8.5\times10^{-5}$
& few$\,\times10^{-7}$ & $<10^{-15}$\\
~$D\to X_u+\gamma$ & & & $\sim10^{-5}$\\
~$D^0\to \rho^0\gamma$ & $1.4\times10^{-4}$& &
$(1-5)\times10^{-6}$\\
~$D^0\to \phi\gamma$ & $2\times10^{-4}$& 
& $(0.1-3.4)\times10^{-5}$\\
\hline
LF or LN Viol.\ 
& & & \\ \hline
~$D^0\to\mu^\pm e^\mp$ & $1.0\times 10^{-4}$& $10^{-7}$ & 0 \\
~$D^+\to\pi^+\mu^\pm e^\mp$ & $3.3\times 10^{-3}$
& few$\times10^{-7}$ & 0 \\
~$D^+\to K^+ \mu^\pm e^\mp$ & $3.4\times 10^{-3}$
& few$\times10^{-7}$ & 0 \\
~$D^+\to \pi^- \mu^+\mu^+$ & $2.2\times 10^{-4}$
& few$\times10^{-7}$ & 0 \\
~$D^+\to K^- \mu^+\mu^+$ & $3.3\times 10^{-4}$
& few$\times10^{-7}$ & 0 \\
~$D^+\to \rho^- \mu^+\mu^+$ & $5.8\times 10^{-4}$
& few$\times10^{-7}$ & 0 \\
\hline
Mixing
& & & \\ \hline
~${}^{^{(}}{\overline {D^0}}{}^{^{)}}\to K^\mp\pi^\pm$ &
$r<0.0037$& $r<10^{-5}$ & \\
& $\Delta M_D<1.3\!\times\!10^{-4}$\,eV & $\Delta M_D<10^{-5}\,$eV &
$10^{-7}$\,eV\\
~${}^{^{(}}{\overline {D^0}}{}^{^{)}}\to K\ell\nu$ & & $r<10^{-5}$ & \\
\hline
\end{tabular}
\end{center}
$^*$ To save space, sources for the measurements and predictions in this table
are not cited here; most may be found in  
Refs.\,\cite{C0_charm} and \,\cite{Strasbourg}.\hfill\break
$\dagger$ at 90\% confidence level.
\vspace{-.1in}
\end{table*}
}

These asymmetries reflect interference due to the 
CKM phase in combination with phase differences from final-state 
interactions.
Experimental evidence suggests substantial final-state effects in
charm decay. For example, the mode $D^0\to K^0 \overline{K^0}$ occurs with a
branching ratio\,\cite{PDG}
\begin{equation} \frac{B(D^0\to K^0 \overline
{K^0})}{B(D^0\to K^+ K^-)} = 0.24\pm 0.09\,, \end{equation} 
even though no spectator diagram can produce this final state, and the two
possible $W$-exchange diagrams
cancel each other (by the GIM mechanism) to good approximation.
This mode could be fed by rescattering of $K^+K^-$ into $K^0\overline{K^0}$.
Large final-state effects are also evident in the case of multibody charm
decays, where Dalitz-plot analyses reveal appreciable phase
differences.\,\cite{E687-Dalitz} These and similar observations underlie the
expectation of $\cal{O}$$(10^{-3})$ direct {\CP} asymmetries in charm.

Additional SM mechanisms for charm {\CP} violation include $K^0$ mixing and 
possible mixing with glueballs or gluonic hybrids. As emphasized by
Xing,\,\cite{Xing} $K^0$ mixing leads to {\em CP} asymmetries of
$\approx$2\,Re$(\epsilon_K)=3.3\times10^{-3}$ in such decays as $D^+\to
K_S\pi^+$ and $D^+\to K_S\ell\nu$. While perhaps not as interesting as direct
charm {\CP} violation, this effect might provide a calibration for systematic
effects in the measurement of small asymmetries. As discussed below, it 
could also represent a unique window into new physics. Close and
Lipkin\,\cite{Close-Lipkin} make the intriguing suggestion that $D$'s could be
mixed with gluonic-hybrid states, with consequent large {\CP}-violating effects.

At present the best limits on direct {\em CP} violation in Cabibbo-suppressed
charm decay come from Fermilab E687\,\cite{Frabetti} and E791\,\cite{E791-CP}
and CLEO\,\cite{Bartelt} (Table~\ref{tab:sens}).
In fixed-target experiments, to correct for the production asymmetry of $D$ \vs
$\overline {D}$, the asymmetry in a Cabibbo-suppressed mode is normalized to
that observed in the corresponding Cabibbo-favored (CFD) mode; this also has
the effect of reducing sensitivity to such systematic effects as trigger,
reconstruction, and particle-identification efficiency differences for
particles \vs antiparticles. In E687 $\approx$10\% sensitivity is achieved.
By extrapolation from E687, the definitive
establishment of a $10^{-3}$ asymmetry requires $_\sim$\llap{$^>$}10$^{9}$
reconstructed
$D$'s, to give $\sim$10$^{7}$ reconstructed charged and (tagged)
neutral $D$'s in SCSD modes.

Although the ratiometric nature of these measurements makes them intrinsically
insensitive to systematic effects, at the sub-$10^{-3}$ level careful attention
will be required to keep systematic uncertainties from dominating.

\subsection{Beyond the Standard Model}

For several reasons, the charm sector is an excellent place to look for {\em
CP} violation arising from physics beyond the Standard Model: 
\begin{itemize}
\item The top-quark loops that in the Standard Model dominate {\CP} violation
in the strange and beauty sectors\,\cite{Rosner} are absent, creating a
low-background window for new physics. 
\item New physics may couple differently
to up-type and down-type quarks\,\cite{Hadeed-Lepto} or couple to quark
mass.\,\cite{Bigi87} 
\item Compared to beauty, the large production cross
sections\,\cite{sigmaD} allow much larger event samples to be acquired, and the
branching ratios to final states of interest are also larger.\,\cite{PDG} 
\item
Many extensions of the Standard Model predict observable effects in charm.
\end{itemize} 
Direct {\CP} violation  in Cabibbo-favored or doubly
Cabibbo-suppressed (DCSD) modes would be a clear signature for new
physics.\,\cite{Burdman,Bigi94} Asymmetries in these as well as in SCSD modes
could reach $\sim$10$^{-2}$ in such scenarios as non-minimal
supersymmetry\,\cite{Bigi94} and in left-right-symmetric
models.\,\cite{left-right,Pakvasa} Bigi has pointed out that a small
new-physics contribution to the DCSD rate could amplify the SM $K^0$-induced
asymmetries to $\cal{O}$$(10^{-2})$ as well.\,\cite{Bigi94}

Many authors have recently emphasized the possibility of observable indirect
{\CP} violation in charm.\,\cite{Bigi94,Wolfenstein}$\,^-$\,\cite{London} 
This of
course depends on charm mixing, which has not been established
experimentally.\,\cite{PDG,E691,E791-mixing} However, the observation of a
wrong-sign signal (which may be mixing, DCSD, or some mixture of the two) at
CLEO\,\cite{Cinabro} has stimulated theorists to consider the large variety of
extensions to the SM in which $D^0$ and $\overline{D^0}$ can display
appreciable {\CP}-violating mixing. These include flavor-changing Higgs
exchange, a fourth generation, $Z$-mediated  FCNC's, 
left-right symmetry, supersymmetry with
quark-squark alignment, leptoquarks, etc.\,\cite{Nir,London} 
At the level discussed in
the literature, such effects are likely to be observable in a
$10^8$-to-$10^9$-charm experiment.

As a specific example, I consider a possible indirect {\CP} asymmetry in $D^0 
(\overline{D^0})\to\pi^+\pi^-$. We would expect $\sim10^7$ tagged $\pi^+\pi^-$
decays per few$\,\times10^9$ reconstructed charm, giving $<10^{-3}$ 
sensitivity. Since here the final state is a {\CP} eigenstate, the {\CP} asymmetry 
is independent of final-state phases and thus directly measures the new-physics 
phase.\,\cite{London}

\section{Other charm physics}

The BTeV/C0 experiment will have unprecedented reach in all areas of charm
physics, including tests of QCD and HQET in meson and baryon spectroscopy and
lifetimes, charm production, Dalitz-plot analyses,
semileptonic form factors, extraction of CKM elements, etc.
Space restrictions preclude further discussion here.

\section{Experimental Apparatus}

Fig.~1 is a sketch of the BTeV/C0 spectrometer as currently conceived. 
The spectrometer is designed for the Tevatron C0 collider interaction hall,
to be upgraded and expanded during the 1998 Main Injector construction period.
It differs from existing collider experiments in that it focuses on heavy-quark 
states produced in the ``forward" direction 
($|\tan{\theta}|_\sim$\llap{$^<$}0.3). This 
approach allows optimal decay-time resolution and is also advantageous for
hadron identification.

The apparatus must have high
interaction-rate capability, large acceptance,
an efficient charm trigger, high-speed and 
high-capacity data acquisition,
good mass and vertex resolution, and good particle 
identification. Of these requirements, the most challenging are the 
trigger and the particle identification. We intend to trigger primarily on the
presence of a decay vertex separated from the primary
vertex.\,\cite{Vertex-trigger}  To reduce occupancy and facilitate vertex
reconstruction at trigger level 1, pixel detectors will be used for vertex
reconstruction.
For efficient, reliable, and compact particle 
identification, we will build a ring-imaging Cherenkov counter.
In other respects the spectrometer will resemble existing large-aperture
fixed-target heavy-quark experiments.

\begin{figure}
\vspace{-1.65in}
\centerline{\hspace{-1.3in}\epsfysize=3.5 in\epsffile{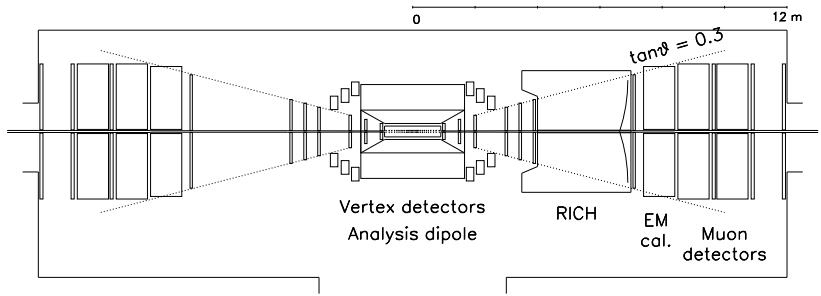}}
\vspace{-.35in}
\caption{Sketch of BTeV/C0 Spectrometer.}
\vspace{-.25in}
\end{figure}

\section{Sensitivity Estimate
\label{sec:sens}}

Our charm sensitivity goal might be achieved in either collider or fixed-target 
mode.  During Tevatron Run II, much early running at C0 may be in fixed-target
mode, \eg using a wire target in the beam halo. We have
estimated\,\cite{C0_charm,Dubna} the fixed-target reconstructed-event yield at
$\approx 1\times10^8$ per $10^7$ seconds of running for an experiment operating
at a 1-MHz interaction rate. Given the higher production cross section,
comparable or greater sensitivity could be available in collider mode even 
with reduced running time.  In addition, we anticipate increasing the
interaction rate beyond 1\,MHz as the Tevatron bunch separation is reduced from
396 to 132\,ns. Ultimately $\approx$5 MHz could be feasible, leading to $>10^9$
charm decays reconstructed.

Whether fixed-target or collider mode is better for charm physics is a detailed
question which probably cannot be answered definitively until data are taken.
For one thing, the forward charm-production cross section has not yet been 
measured in 2\,TeV $p\bar p$ collisions. There are also subtleties, for
example biases in mixing studies may arise from $b\to c$ cascade decays. These
would be suppressed by two orders of magnitude in fixed-target relative to
collider mode, due to the reduced beauty production cross section.

\section{Conclusions}

A hadroproduction experiment 
capable of reconstructing $>10^9$ charm events is feasible using detector,
trigger, and data acquisition technologies that exist or are under
development. Such an experiment could observe
direct {\em CP} violation in charm decay at the level expected in the Standard
Model and substanially extend the discovery reach for new physics.

\end{document}